\documentclass[paper]{agujournal2019}
\usepackage{url}
\usepackage{lineno}
\usepackage[inline]{trackchanges} 
\usepackage{soul}
\usepackage{color}
\usepackage{amsmath}

\addeditor{VG}
\addeditor{AD}
\addeditor{AA}

\draftfalse

\newcommand{\planss} {{Planetary Space Science }}  
\newcommand{\ssr}{   {Space Sci. Rev. }}

\newcommand{\jgr}{   {J. Geophys. Res.}}
\newcommand{\grl}{   {Geophys. Res. Lett.}}

\newcommand{\apjl}{   {Astrophys. J. Lett.}}

\graphicspath{{./}{figures/}}

\journalname{Geophysical Research Letters}

\begin{document}


\title{Relativistic electron precipitation by EMIC waves: importance of nonlinear resonant effects}

\authors{Veronika S. Grach \affil{1}, Anton V. Artemyev \affil{2,3}, Andrei G. Demekhov \affil{1,4},
Xiao-Jia Zhang \affil{2}, Jacob Bortnik \affil{2},  Vassilis Angelopoulos \affil{2}, R. Nakamura\affil{5},  E. Tsai\affil{2}, C. Wilkins\affil{2}, O. W. Roberts\affil{5}}
\affiliation{1}{Institute of Applied Physics, Russian Academy of Sciences, Nizhny Novgorod, Russia}
\affiliation{2}{University of California, Los Angeles, Los Angeles, USA}
\affiliation{3}{Space Research Institute RAS, Moscow, Russia.}
\affiliation{4}{Polar Geophysical Institute, Apatity, Russia}
\affiliation{5}{Space Research Institute, Austrian Academy of Science, Graz, Austria}

\correspondingauthor{Veronika S. Grach}{vsgrach@ipfran.ru}

\begin{keypoints}
\item ELFIN observations of EMIC-driven precipitation of relativistic electrons
\item Test-particle model reproduce precipitating electron energy range
\item Nonlinear resonant effects cannot stop the diffusive scattering into loss-cone
\end{keypoints}

\begin{abstract}
Relativistic electron losses in Earth’s radiation belts are usually attributed to electron resonant scattering by electromagnetic waves. One of the most important wave mode for such scattering is the electromagnetic ion cyclotron (EMIC) mode. Within the quasi-linear diffusion framework, the cyclotron resonance of relativistic electrons with EMIC waves results in very fast electron precipitation to the atmosphere. However, wave intensities often exceed the threshold for nonlinear resonant interaction, and such intense EMIC waves have been shown to transport electrons away from the loss cone due to the {\it force bunching} effect. In this study we investigate if this transport can block electron precipitation. We combine test particle simulations, low-altitude ELFIN observations of EMIC-driven electron precipitation, and ground-based EMIC observations. Comparing simulations and observations, we show that, despite of the low pitch-angle electrons being transported away from the loss cone, the scattering at higher pitch angles results in the loss cone filling and electron precipitation. 
\end{abstract}

\section{Introduction}\label{sec:intro}
Dynamics of relativistic electron fluxes in the Earth's inner magnetosphere is largely controlled by the competition of acceleration processes \cite<due to resonances with whistler-mode waves and/or radial transport, see, e.g.,>[and references therein]{Shprits08:JASTP_local,Shprits08:JASTP_transport,Millan&Baker12} and losses, which are significantly contributed by electron resonant interaction with electromagnetic ion cyclotron (EMIC) waves \cite<see, e.g.,>[]{Usanova14,Ma15,Drozdov17}. EMIC waves are generally much stronger than whistler-mode waves, and the resulted relativistic electron scattering can be incredibly fast \cite<see estimates of EMIC-driven diffusion rates in>[]{Summers&Thorne03,Kersten14,Ni15,Shprits16}. The main approach to quantify the EMIC-driven electron losses in the radiation belt models is the quasi-linear theory \cite<e.g.,>[]{Thorne&Kennel71,Albert03,Glauert&Horne05}, which operates with diffusion rates calculated under the assumption of weak wave intensity \cite{Vedenov62,Drummond&Pines62,Andronov&Trakhtengerts64,Kennel&Petschek66}. Spacecraft observations, however, often show EMIC wave intensities sufficiently large to exceed the threshold of quasi-linear diffusion applicability \cite<see discussion in>[]{Wang17:emic,Grach21:emic}. Therefore, an important question in modeling relativistic electron losses is how does electron scattering change for intense EMIC waves? 

In contrast to diffusion by low amplitude waves, intense waves resonate with electron nonlinearly. Such nonlinear resonant interaction includes phase trapping and phase bunching, which provide direct transport of electrons in pitch-angle space \cite<see reviews by>[]{Karpman74:ssr,Omura91:review,Shklyar09:review,Albert13:AGU,Artemyev18:cnsns}. For relativistic electron resonances with EMIC waves, phase bunching results in electron transport away from the loss cone \cite{Albert&Bortnik09}, whereas phase trapping quickly moves electrons into the loss cone and may result in strong precipitation \cite{Omura&Zhao13,Kubota15,Kubota&Omura17}. Moreover, in low pitch angles (close to the loss cone), the nonlinear wave-particle interaction is significantly modified \cite<see, e.g.,>[]{Kitahara&Katoh19,Albert21,Artemyev21:pop}, which is particularly important for relativistic electron resonances with EMIC waves but not included in most existing models.
 Around the loss cone, EMIC wave interaction with electrons does not include phase trapping but is dominated by the so-called force bunching effect \cite{Grach&Demekhov20}, which leads to rapid pitch-angle increases. Such direct transport away from the loss cone may block the electron precipitation \cite<see discussion in>[]{Bortnik22}. Thus, it remains paradoxical where increased EMIC wave amplitude above the threshold of nonlinear resonant interaction may result in decreased efficiency of electron scattering into loss cone. 

In this case study we focus on detailed investigation of relativistic electron scattering by intense EMIC waves. We use electron precipitation observations at low-altitude ELFIN CubeSats \cite{Angelopoulos20:elfin} and conjugate EMIC wave observations at the Lovozero ground-based station \cite{Fedorenko14}. We also use measurements from the Magnetospheric Multiscale (MMS) mission \cite{Burch16} to probe the equatorial plasma conditions important for electron resonances with EMIC waves \cite<see>{Summers05,Summers07:rates}. Spacecraft and ground-based measurements are combined with test particle simulations that include realistic models of EMIC wave propagation \cite<see>{Grach21:emic}. We show that the force phase bunching indeed quickly moves relativistic electrons to higher pitch angles, but EMIC waves also scatter high pitch-angle electrons directly into loss cone. The competition of these two processes results in loss-cone filling and electron precipitation observed by ELFIN. The paper is structured as follows: descriptions of spacecraft and ground-based observations in Section~\ref{sec:observations}, discussion of test particle simulation results in Section~\ref{sec:simulation}, and conclusions in Section~\ref{sec:conclusions}.

\section{Spacecraft observations}\label{sec:observations}
\begin{figure*}
\centering
\includegraphics[width=1\textwidth]{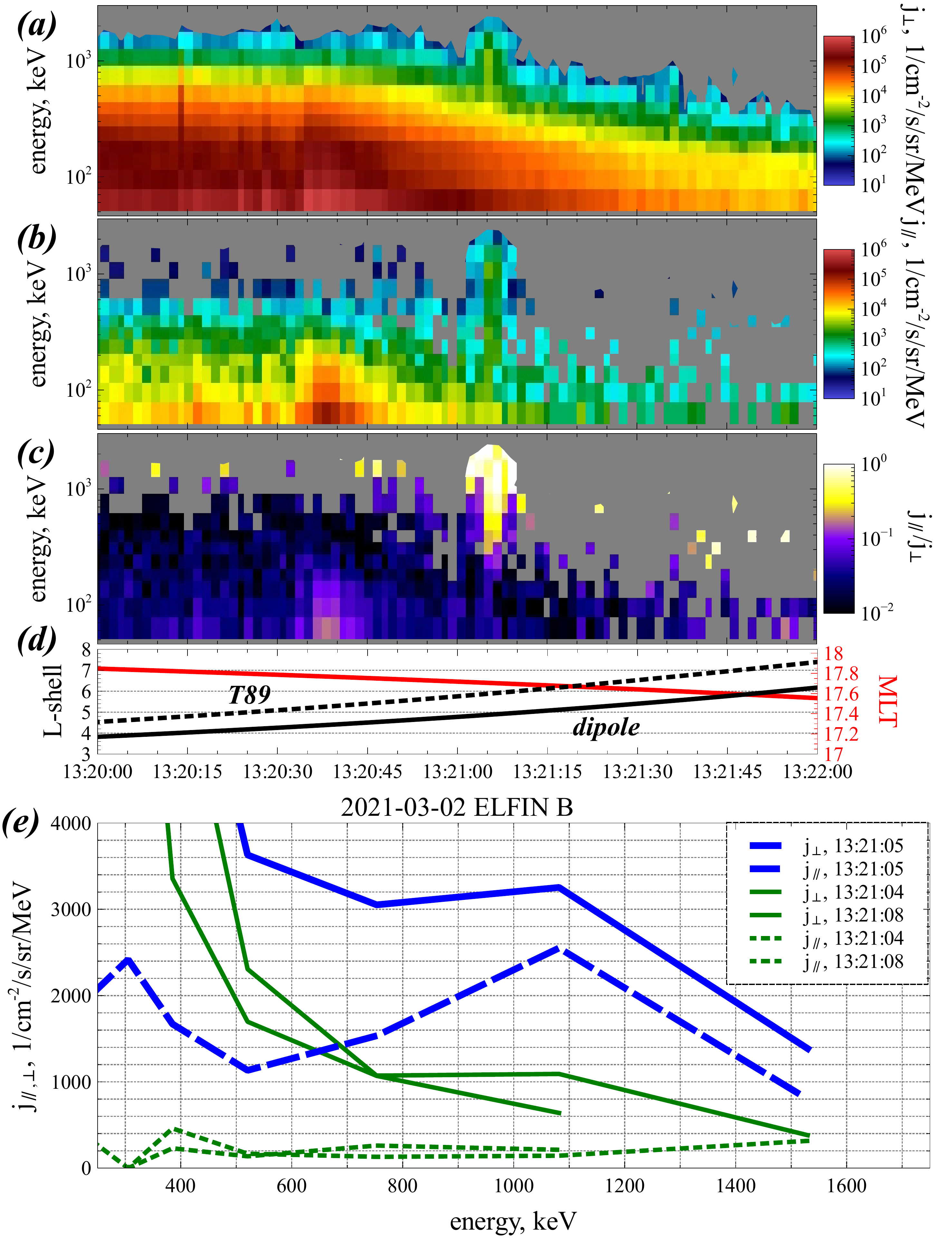}
\caption{An overview of ELFIN B observations on 2 March 2021: energy spectrum of trapped electron fluxes (a), spectrum of precipitating electron fluxes (b), precipitating-to-trapped flux ratio (c), ELFIN B MLT and $L$-shell calculated with the dipole and \cite{Tsyganenko89} magnetic field models (d), spectra of trapped ($\perp$) and precipitating ($\parallel$) fluxes around the precipitation burst (e).
\label{fig1}}
\end{figure*}

Figure \ref{fig1} shows the event overview with ELFIN observations of EMIC-driven electron precipitation. Energetic particle detector (EPDe) onboard ELFIN measures electrons in the $[50,7000]$ keV energy range (16 channels, $\Delta E/E\approx0.4$) and $\in[0,360^\circ]$ pitch-angle range (16 sectors, $\Delta\alpha=22.5^\circ$) with a 1.5~s (half spin) resolution \cite{Angelopoulos20:elfin}. Panel (a) shows the energy spectrum of trapped electrons (pitch angle $\alpha\in[\alpha_{LC},180^\circ-\alpha_{LC}]$, where $\alpha_{LC}$ is the local loss-cone angle). ELFIN moved from low $L$-shells to higher $L$-shells on the dusk flank (see panel (d)), along which ELFIN traversed the plasmasphere (according to equatorial spacecraft measurements, the plasmapause is located around $L(T89)\sim 6$, see discussion below) and outer radiation belt. Around the plasmapause, at 13:21:05 UT, trapped electron fluxes of relativistic energies ($>700$ keV) show a local maximum. Note that the locally trapped electrons at ELFIN map to small equatorial pitch angles, right outside the equatorial loss cone. Such transient flux increases shall be interpreted as a spatially localized enhancement of small pitch-angle electron fluxes at the equator. Panels (b,c) confirm that increases of trapped fluxes are accompanied by strong precipitation of relativistic electrons. There are two important features of this transient precipitation: First, precipitating fluxes and ratio of precipitating-to-trapped fluxes maximize at relativistic energies ($>700$ keV) without a significant precipitation of $<500$ keV electrons, which would indicate electron scattering by whistler-mode waves (compare the precipitation at 13:21:05 UT and precipitation likely driven by plasmaspheric hiss at 13:20:35-13:20:45 UT; also see the discussion of ELFIN observations of whistler-mode wave driven precipitation in \citeA{Mourenas21:jgr:ELFIN,Artemyev21:jgr:ducts}); Second, precipitating fluxes at relativistic energies (around $\sim 1$~MeV) almost reach the strong diffusion limit (see the comparison of precipitating and trapped electron spectra in panel (e)). These two features strongly suggest that the observed precipitation is driven by electron scattering by EMIC waves \cite<see similar equatorial observations of small pitch-angle flux enhancements associated with EMIC waves in>[]{Zhu20:emic}, which resonate with relativistic electrons most effectively \cite{Albert03,Kersten14,Ni15,Shprits17} and are well able to cause precipitation in the strong diffusion limit \cite{Omura&Zhao13,Kubota&Omura17,Grach&Demekhov20}. Panels (b,e) show that in addition to the clear precipitation peak at $\sim 1$~MeV, there exists moderate precipitation down to $50$~keV. Such weak electron precipitation at lower energies may be attributed to electron scattering by whistler-mode hiss waves around the plasmasphere \cite<see discussion of hiss wave localization around the plasmasphere in, e.g.,>[]{Malaspina20:hiss}.

\begin{figure*}
\centering
\includegraphics[width=1\textwidth]{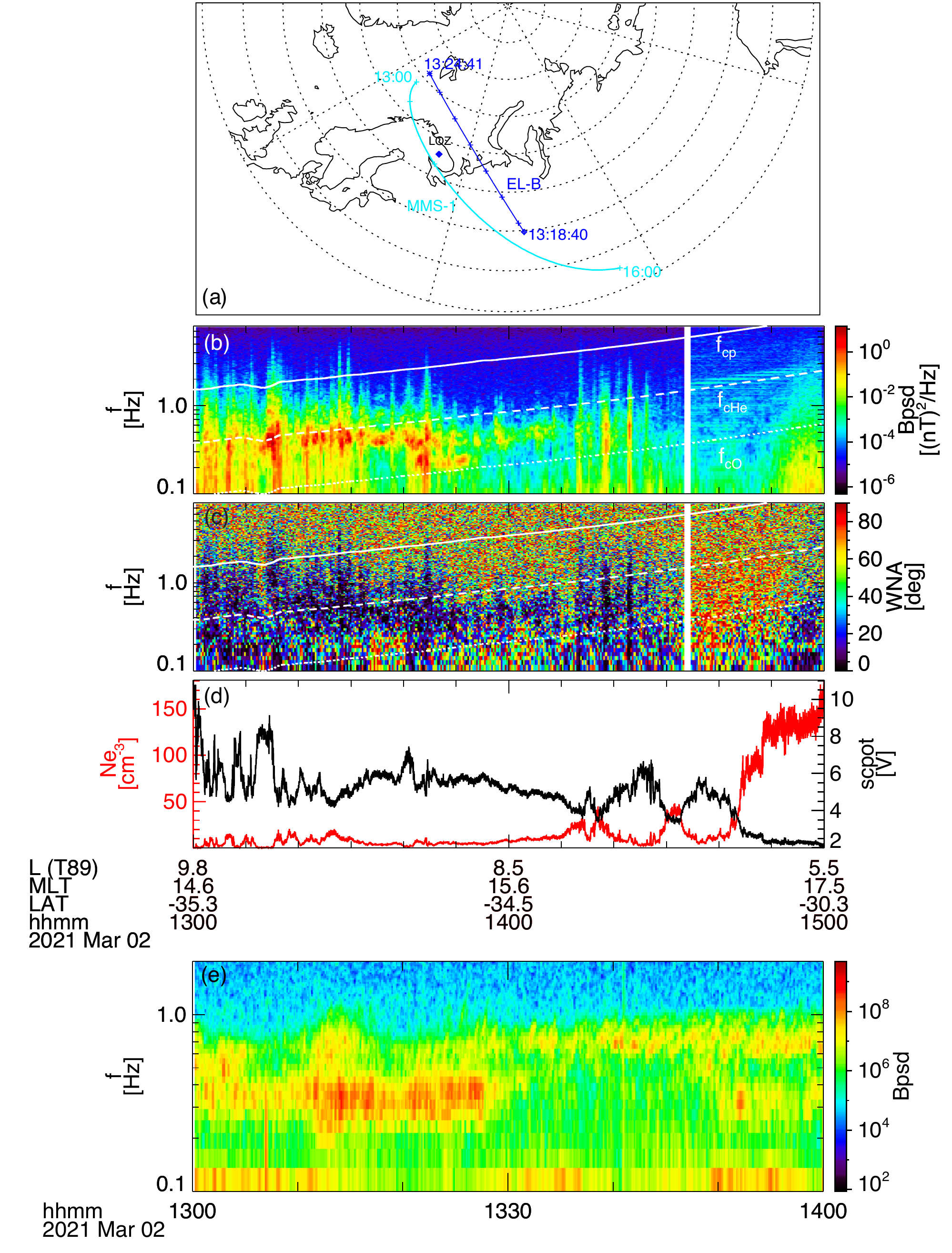}
\caption{Overview of EMIC wave measurements: projections of ELFIN B and MMS \#1 orbits to the northern hemisphere with the ground-based LOZ station (a), MMS magnetic field spectrum in the EMIC wave frequency range (b), EMIC wave normal angle distributions (c), MMS spacecraft potential and the inferred density (d), spectrum of EMIC wave magnetic field measured at the LOZ station (e).
\label{fig2}}
\end{figure*}

Figure \ref{fig2}(a) shows that the ELFIN orbit with EMIC-driven electron precipitation is projected nearby the ground-based magnetometer station (LOZ), whereas the same ($MLT$, $L$-shell) region was traversed by MMS spacecraft $\sim 1.5$ hours later. Due to the time difference between MMS and ELFIN crossings of $L\in [6,7]$, field-aligned helium band EMIC waves detected by MMS (see panels (b,c)) cannot be directly associated with the precipitation at ELFIN \cite<although the EMIC wave source region may survive for several hours, see>[]{Engebretson15}. Note that MMS were at middle latitudes, well below the equator, and hence cannot detect all near-equatorial EMIC waves that possibly persist after $\sim$ 14:00 UT.

We use spacecraft potential measurements \cite{Ergun16:ssr,Lindqvist16,Torkar14} and the inner magnetosphere temperature model \cite{Boardsen14} to estimate the cold plasma density at MMS \cite{Andriopoulou16, Andriopoulou18}. Figure \ref{fig2}(d) shows plasma density enhancements around the plasmapause, at $L\in [8,6]$, with the peak density reaching $\sim 40$~cm$^{-3}$. Such density enhancements can survive for hours \cite<e.g.,>[]{Goldstein14}, which can then guide EMIC wave propagation to the ionosphere. Indeed, the LOZ station detected frequency-banded, $\in [0.2,1]$ Hz, waves resembling properties of EMIC emissions measured by MMS (see panel (e)). More importantly, ground-based observations of EMIC waves are in a good conjunction with ELFIN precipitation measurements. Therefore, we may assume that the precipitation at ELFIN is driven by helium band EMIC waves, whereas localized plasma density peaks reduce the electron resonant energies to $\sim 1$ MeV, as shown in the precipitation energy at ELFIN.

\section{Simulation setup and results}
\label{sec:simulation}
To investigate the mechanism of electron precipitation observed by ELFIN, we perform a test particle simulation with observed EMIC wave characteristics. We use the dipole model of geomagnetic field $\mathbf{B}_0$ and gyrotropic model of plasma density $N= N_{eq}B_0/B_{0eq}$, where the subscript "$eq$" denotes equatorial values. Based on MMS measurements (see Fig. \ref{fig2}(d)), we choose $L=6.2$, $N_{eq}=40$~cm$^{-3}$, and use the plasmaspheric ion composition  $N_{\rm{H}^+}=0.8N$, $N_{\rm{He}^+}=0.2N$ from \citeA{Horwitz90:plasmasphere}. 
 Wave packet parameters (constant wave frequency $f=0.95f_{\rm{He}^+}=0.48$~Hz and duration $\tau=120$~s) are set in agreement with ground-based (LOZ) observations;  here $f_{\rm{He}^+}$ is helium gyrofrequency. We also assume a Gaussian amplitude profile with the maximum wave amplitude $B_{\rm{wm}}=2$~nT. 

\begin{figure} \renewcommand{\baselinestretch}{1.0}
\begin{center}
  \centering
\includegraphics[width=\textwidth]{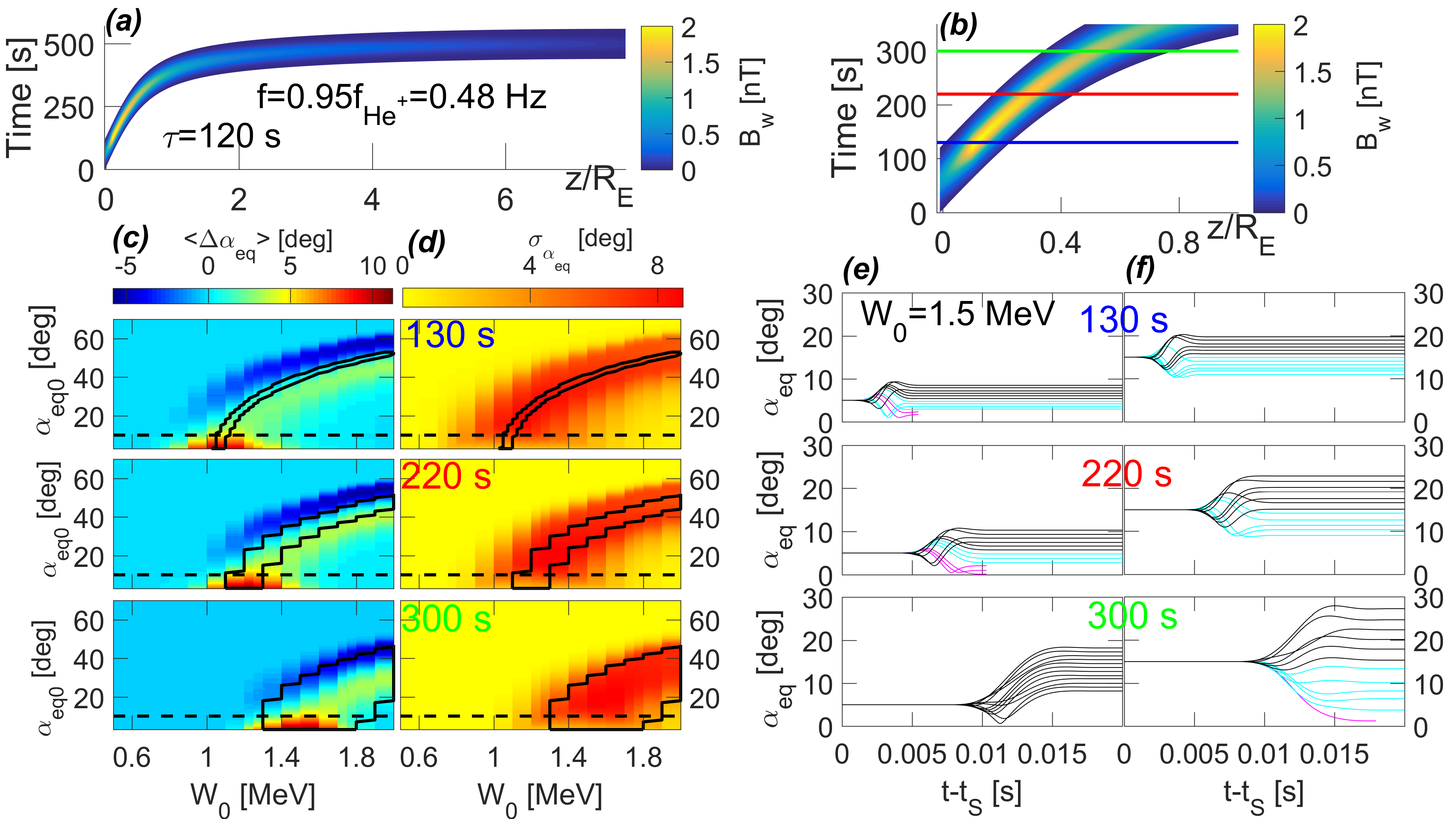}
\caption{(a-b): The spatial-temporal distribution of EMIC wave field as the wave packet propagates away from its source region near the equator (z=0). Horizontal lines in (b) mark the start times ($t_S$) of electron trajectories in three runs. (c-d): Phase averaged change in $\alpha_{eq}$ and rms $\sigma_{\alpha_{eq}}$ after a single pass through the wave packet at times $t_S$  as a function of initial energy $W_0$ and equatorial pitch angle $\alpha_{eq0}$. Black contours show unperturbed resonant domains (where the exact resonance condition with $\Delta=0$ is satisfied); dashed lines correspond to $\alpha_{eq0}=10^\circ$. (e-f): Pitch-angle trajectories for particles with $W_0=1.5$~MeV and $\alpha_{eq0}=5^\circ; 15^\circ$. Black, cyan, and magenta lines correspond, respectively, to those with pitch-angle increasing and decreasing, and particles scattered into the loss cone.}
  \label{Fig_Packets_DTH_Traj}
     \end{center}
\end{figure}

EMIC wave packet is generated in a small region near the equator and then propagates along the magnetic field lines. The spatial-temporal distribution of the wave field is retrieved using the ray-tracing technique, as shown in Figure \ref{Fig_Packets_DTH_Traj}(a) (see also the distribution zoomed around the equator in panel (b)). In ray-tracing, the wave energy flux $\mathcal{E}\propto B_\mathrm{w}^2 V_{\mathrm{gr}}/(8\pi)$ for each point of the packet remains constant, and the group velocity $V_{\mathrm{gr}}$ corresponds to the local wave dispersion relation \cite<see details of wave model in>{Grach21:emic}. 

The model wave field is then used to evaluate trajectories of test electrons interacting with EMIC waves at the anomalous cyclotron resonance. We use the gyro-averaged electron equations of motion \cite<see details in>[]{Grach&Demekhov20, Grach21:emic}:
 \begin{linenomath}
\begin{equation}
 \label{eqW}
    \frac{\mathrm{d} W}{\mathrm{d}t}= - e v_\bot |E_\mathrm{w}| \sin\Psi;
    \end{equation}
    \end{linenomath}
     \begin{linenomath}
\begin{equation}
 \label{eqIbot}
    \frac{\mathrm {d}{I_\bot}}{\mathrm{d}t}=
-\frac{2e}{mB_0}p_\bot(1-n_{||}\beta_{||})|E_\mathrm{w}|\sin\Psi;
\end{equation}
    \end{linenomath}
       \begin{linenomath}
\begin{equation}
      \label{eqPsi}
\frac{\mathrm{d} \Psi}{\mathrm{d}t} =-\Delta
-\frac{e}{p_{\bot}} (1-n_{||}\beta_{||})|E_\mathrm{w}|\cos\Psi;
 \end{equation}
    \end{linenomath}
       \begin{linenomath}
\begin{equation}
  \label{eqZ}
 \frac{\mathrm{d}z}{\mathrm{d}t}=\frac{p_{||}}{m\gamma}.
 \end{equation}
    \end{linenomath}
Here subscripts ${||}$ and $\bot$ denote projections to the parallel and transverse directions with respect to $\mathbf{B}_0$; $W=(\gamma-1)mc^2$ and $I_\bot = {p_\bot^2}/({mB_0})$ are the electron kinetic energy and the first adiabatic invariant; $\gamma=\sqrt{1+\left[{p}/({mc})\right]^2}$ is the Lorentz factor; $m$, $p$ and $v$ are the electron rest mass, momentum, and velocity; $\beta_{||}=v_{||}/c$, $n_{||}=kc/\omega$; $\omega=2\pi f$, $k$ and 
$E_\mathrm{w}$ are wave frequency, wave number, and slowly changing wave electric field amplitude; $z$ is the coordinate along the geomagnetic field with $z=0$ corresponding to the equator; $\Psi=\vartheta-\varphi$, $\varphi$ is the gyrophase in the geomagnetic field $\mathbf{B}_0$; $\vartheta={\omega}/{c}\int n_{||}\mathrm{d}z - \int \omega \mathrm{d}t$ is the wave phase, $\Delta=\omega-kv_{||}+\Omega_\mathrm{c}/\gamma$ is the mismatch from the resonance; $\Omega_\mathrm{c}=eB_{0}/mc$, with $e>0$ being the elementary charge.

These equations of motion (\ref{eqW})--(\ref{eqZ}) are solved numerically by Bogacky-Shampine variant of the Runge-Kutta method. We use $\sim 0.5\cdot10^6$ trajectories with $19$ initial energies $W_0\in[0.5,2.0]$~MeV, $135$ initial equatorial pitch angles $\alpha_{eq0}\in[3^\circ,70^\circ]$, and $180$ initial phases uniformly distributed in $[0, 2\pi)$. Note that for the considered $L$, the loss-cone angle is $\alpha_{eq}^{LC} \approx 2.7^\circ$. To investigate variations of electron scattering as the wave packet evolves along magnetic field lines (see Fig. \ref{Fig_Packets_DTH_Traj}(a,b)), we run three simulations where the electron trajectories start from the equator at times $t_S=130, 220$, and $300$~s, respectively. Electron energy remains nearly constant during the resonant interaction with EMIC waves (because the wave frequency is much smaller than electron gyrofrequency), and hence we characterize the resonant interaction by the change of equatorial pitch angles, $\alpha_{eq}$. We calculate the mean electron pitch angle change $\langle\Delta\alpha_{eq}\rangle$ and spread $\sigma_\alpha=\sqrt{\langle(\Delta\alpha_{eq})^2\rangle-\langle\Delta\alpha_{eq}\rangle^2}$ as functions of initial energy and pitch angle.

Figure \ref{Fig_Packets_DTH_Traj}(c,d) shows $\langle\Delta\alpha_{eq}\rangle$, $\sigma_\alpha$ maps for three different test particle runs (i.e., three different wave field distributions at the resonant latitudes). Electrons with small initial pitch angles show a significant $\langle\Delta\alpha_{eq}\rangle>0$ at the resonant energy that increases with time (with the resonant latitudes where the wave-packet intensity maximises) from 1~MeV to 1.5~MeV (see red regions in Figure \ref{Fig_Packets_DTH_Traj}(c)). Such a pitch-angle {\it reflection} from the loss cone has been found by \cite{Lundin&Shkliar77,Inan78} and described for EMIC waves in \citeA{Grach&Demekhov20} as {\it force phase bunching}. This pitch-angle change is described by the second term of the right-hand side of Eq. (\ref{eqPsi}): the direct influence of Lorentz force on the particle phase is strong for small $p_\perp$ and can exceed the inertial (or kinematic) bunching effect of the the first term in Eq. (\ref{eqPsi}). Figure \ref{Fig_Packets_DTH_Traj}(e) shows this force phase bunching with pitch-angle increase in test trajectories \cite<see also statistical investigation of this effect in>{Bortnik22}.

For electrons with pitch angles exceeding $\sim 10^\circ$, the resonant interaction with EMIC waves results in predominantly $\langle\Delta\alpha_{eq}\rangle<0$ (see blue stripes in Fig. \ref{Fig_Packets_DTH_Traj}(c)). However, this electron drift toward the loss cone is not very strong ($\langle\Delta\alpha_{eq}\rangle>-5^\circ$), because the corresponding initial energies/pitch angles are outside the domain of resonant interactions (shown by black curves). For $\alpha_{eq}>10^\circ$ within the resonant domain, the pitch-angle drift is almost absent, $\langle\Delta\alpha_{eq}\rangle\approx 0$. Spread in pitch-angle scattering, however, is quite large for $\alpha_{eq}>10^\circ$ electrons: $\sigma_\alpha>5^\circ$ (see Fig. \ref{Fig_Packets_DTH_Traj}(d)). Several test particle trajectories in Fig. \ref{Fig_Packets_DTH_Traj}(f) shows that such diffusive scattering (almost symmetric in the number of particles with $\Delta\alpha_{eq}>0$ and $\Delta\alpha_{eq}<0$) may move $10^\circ$--$15^\circ$ electrons directly into the loss cone, and these electrons should move to the atmosphere after a single resonant interaction. Note that in the considered case of relatively strong wave amplitude and short packet duration, the resonant interaction zone is determined by the packet length (see Figures \ref{Fig_Packets_DTH_Traj}3e,f and resonant range in Figures \ref{Fig_Packets_DTH_Traj}c,d). This situation is quite common for realistic EMIC wave packets \cite{Grach&Demekhov20, Grach21:emic}.
    
\begin{figure} \renewcommand{\baselinestretch}{1.0}
\begin{center}
  \centering
\includegraphics[width=\textwidth]{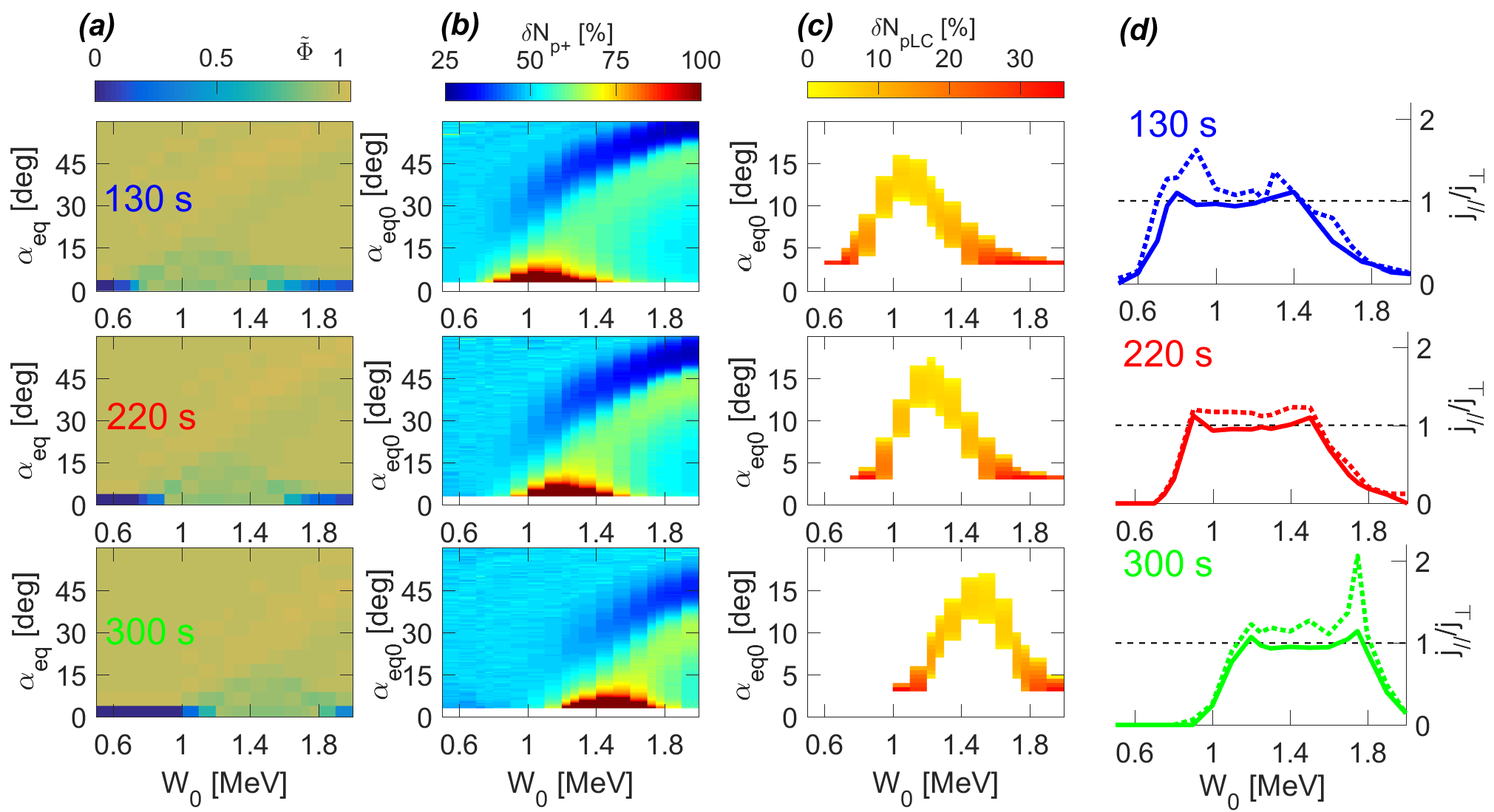}
\caption{Distribution of the particle ensemble after a single pass through the wave packet. Top, middle, and bottom rows correspond to start times of $130$, $220$, and $300$ s, respectively. (a): Pitch-angle distribution; 
(b) and (c): the fraction of particles with increased pitch angles and scattered into the loss cone, respectively; (d) the ratio between the precipitating flux and the flux of trapped particles with $\alpha_{eq}^{LC} \leq \alpha_{eq}< \alpha_{eq}^{LC} +5^\circ$ (solid line). Dotted lines in column (d) show the maximum ratio of the precipitating and trapped fluxes during the 5-second simulation.}
  \label{Fig_Distr_Fluxes}
     \end{center}
\end{figure}

To investigate effects of electron scattering by EMIC waves on electron dynamics, we follow the approach proposed in \cite{Grach&Demekhov20, Grach21:emic}. Assigning each test particle a specific weight, we specify an initial pitch angle distribution $\Phi_{\alpha_{eq}}^{(0)}$ (uniform in our case) and calculate the distribution function $\Phi_{\alpha_{eq}}$ after the wave-particle interactions. The normalized distribution $\tilde{\Phi}=\Phi_{\alpha_{eq}}/\Phi_{\alpha_{eq}}^{(0)}$ is shown in Figure \ref{Fig_Distr_Fluxes}a. The loss cone (the smallest pitch-angle bin) is empty at the beginning, but becomes filled by electrons scattered by EMIC waves. Note that the energy range of the loss-cone filling changes with time because different energies resonate with EMIC wave-packet at different latitudes (at different times). 

Figure \ref{Fig_Distr_Fluxes}b shows the percentage of particles (in each pitch-angle/energy bin) that increase their pitch angles during the simulation. This distribution mainly repeats $\langle\Delta\alpha_{eq}\rangle$ distribution from Fig. \ref{Fig_Packets_DTH_Traj}c. Almost all resonant electrons with small initial pitch angles move away from the loss cone ($\delta N_{p+}\approx 100$\%). Resonant electrons with pitch angles above $10^\circ$ are scattered in a diffusive way: about half electron populations increase pitch angles ($\delta N_{p+}\approx 50$\%) and the other half decrease pitch angles.

Figure \ref{Fig_Distr_Fluxes}c shows distribution of percentage of electrons that are scattered directly into the loss cone after a single resonant interaction with EMIC waves. For the main energy range of resonance, a population of $>5^\circ$ electrons ($\sim 10$\% of total electron population) are scattered into the loss cone. Such electrons have initial pitch angles sufficiently large to avoid the positive drift away from the loss cone (compare Figs. \ref{Fig_Packets_DTH_Traj}c and \ref{Fig_Distr_Fluxes}c), i.e., electrons can jump over the small pitch-angle range with $\langle\Delta\alpha_{eq}\rangle>0$ and move directly into the loss cone.

To compare simulation results with ELFIN data, we plot the ratio of the precipitating flux and flux of trapped particles with equatorial pitch angles of $[\alpha_{eq}^{LC}, \alpha_{eq}^{LC}+5^\circ]$. Figure \ref{Fig_Distr_Fluxes}d shows this ratio (solid curve) as a function of energy for three time moments. The energy range $[0.8,1.8]$ MeV corresponds to very strong precipitation on the strong diffusion limit (with precipitating fluxes reaching the trapped flux level). Comparison of Fig. \ref{Fig_Distr_Fluxes}d and Figs. \ref{fig1}c,e confirms that simulation reproduces the observed energies of strong precipitation. 

We also analyzed the effects of long-term interactions, for which we run three simulations in the time interval of $[t_S-2.5~\mbox{s}, t_S+2.5~\mbox{s}]$, corresponding to $6-12$ bounce oscillations for electrons with energies of $0.5$ to $2$~MeV. On this time scale, wave packet evolution is insignificant, so temporal dynamics is determined by the particle redistribution over pitch angles. Time averaged precipitating flux is roughly the same as in the case of the single interaction, but there are also transient bursts of precipitation deviating from the averaged level. For illustration, the maximum ratio $j_{||}/j_\perp$ is shown in Figure  \ref{Fig_Distr_Fluxes}d as dotted lines. One can see that in some cases precipitating flux can be twice as high as the trapped flux. 

\section{Discussion and Conclusions}
\label{sec:conclusions}
This study is focused on a detailed investigation of EMIC-driven relativistic electron precipitation as observed by ELFIN CubeSats and reproduced in numerical simulations. We aim to examine the effect of electron force phase bunching that moves small pitch-angle electrons away from the loss cone and may potentially block their precipitation \cite<see>[]{Bortnik22}. Observationally driven test particle simulations show that although the force phase bunching indeed prevents precipitation of $\alpha_{eq}<10^\circ$ electrons, electrons with $\alpha_{eq}>10^\circ$ can still be efficiently scattered towards small pitch angles. Such a scattering may move $\alpha_{eq}>10^\circ$ electrons directly into the loss cone, and these electrons will be precipitated without having a chance to be transported away from the loss cone by the force phase bunching effect. Precipitating electron measurements at low altitudes by ELFIN confirm the strong precipitation within the energy range of expectation from simulations. 

Therefore, the main conclusion of this study is that the force phase bunching cannot completely prevent precipitation, but may change the pitch-angle range of precipitating electrons. This nonlinear effect should influence the electron pitch-angle distributions in association with EMIC waves, as measured by near-equatorial spacecraft \cite<e.g.,>[]{Bingley19, Zhu20:emic}, but can unlikely change the strong precipitating rate provided by EMIC waves in radiation belt models. Therefore, the force phase bunching must be taken into account for short-term simulations and interpretation of observations of EMIC-driven electron precipitation bursts, but may have a secondary importance for long-term simulations of radiation belt dynamics.

\acknowledgments
We are grateful to NASA's CubeSat Launch Initiative for ELFIN's successful launch in the desired orbits. We acknowledge early support of ELFIN project by the AFOSR, under its University Nanosat Program, UNP-8 project, contract FA9453-12-D-0285, and by the California Space Grant program. We acknowledge critical contributions of numerous volunteer ELFIN team student members. We appreciate the access to LOZ Pc1 data (PI: Yury Fedorenko). X.-J. Z. acknowledges support from the NSF grant 2021749. V.S.G. acknowledges support from RSF grant 19--72--10111 (numerical simulations in Section 3). A.V.A, X.-J.Z., and J.B. acknowledge support from the NASA grant 80NSSC20K1270. Analysis of the spacecraft potential data at IWF (R.N. and O. W. R.) is supported by Austrian FFG Project No. ASAP15/873685

\section*{Open Research} \noindent 
ELFIN data is available at https://data.elfin.ucla.edu/.\\ \noindent 
MMS data is available at https://lasp.colorado.edu/mms.\\ \noindent 
LOZ data is available at http://aurora.pgia.ru:8071/. \\ \noindent 
Data analysis was done using SPEDAS V4.1 \cite{Angelopoulos19} available at https://spedas.org/.



\end{document}